\documentclass[prl,twocolumn]{revtex4-2}
\usepackage{graphicx}
\usepackage{amssymb,amsmath,stmaryrd,tabularx}
\usepackage{bm}
\usepackage{enumerate}
\usepackage{forest}
\newcommand{\NBI}{Niels Bohr Institute, University of Copenhagen, Blegdamsvej 17, DK-2200 Copenhagen, Denmark.}
\usepackage[utf8]{inputenc}
\usepackage[english]{babel}
\usepackage{booktabs}
\usepackage{csquotes}
\usepackage{amsthm}
\usepackage{amsfonts}
\usepackage{mathtools}
\usepackage[version=4]{mhchem}
\usepackage{color}  
\usepackage{hyperref}
\usepackage{listings}
\usepackage{cleveref}
\usepackage{todo}
\usepackage{nth}
\usepackage{siunitx}

\newcommand{\gv}[1]{\ensuremath{\mbox{\boldmath$ #1 $}}}

\begin{document}
\title{Enhanced dispersion in intermittent multiphase flow}

\author{Joachim Mathiesen$^{1,2}$, Gaute Linga$^{2}$, Marek Misztal$^{1}$, Fran\c{c}ois Renard$^{2,3}$ and Tanguy Le Borgne$^{2,4}$}
\affiliation{$^1$\NBI\\$^2$The Njord Centre, Departments of Geosciences and Physics, University of Oslo, 0316 Oslo, Norway.\\ $^3$ISTerre, University Grenoble Alpes, University Savoie Mont Blanc, CNRS, IRD, 38000 Grenoble, France\\ $^4$Geosciences Rennes, UMR 6118, University Rennes 1, CNRS, 35042 Rennes, France}

\begin{abstract}
Transport in multiphase flow through porous media plays a central role in many biological, geological, and engineered systems. 
Here, we use numerical simulations of transport in immiscible two-phase flow to investigate dispersion in multiphase porous media flow. While dispersion in the main flow direction is similar to that of a single-phase flow and is governed by the porous media structure, we find that transverse dispersion exhibits fundamentally different dynamics. The repeated activation and deactivation of different flow pathways under the effect of capillary forces lead to intermittent flow patterns, strongly enhancing dispersion. We show that the transverse dispersion is controlled by the length scale of fluid clusters, and thus inversely proportional to the square root of the Bond number, the ratio of the force driving the flow and the surface tension. This result leads to a new scaling law relating multiphase flow dynamics to transport properties, opening a range of fundamental and engineering applications.
\end{abstract}

\maketitle

The transport, dispersion, and mixing of solutes are key ingredients in the evolution of a host of natural and engineered porous media \citep{cushman2013physics, dentz2011mixing}. In single phase flow, the structure of the pore space generally results in a broad distribution of flow velocities, leading to non-Fickian dispersion dynamics \citep{lowe1996hydrodynamic, kandhai2002influence, neuman2009perspective,bijeljic2011signature, misztal2015simulating} often described in terms of Continuous Time Random Walk approaches \citep{berkowitz2006modeling, leborgne2008lagrangian, fouxon2016solvable,  deanna2013flow, holzner2015intermittent, kang2014pore}.
In porous media, the presence of several fluid phases, like in unsaturated soils, geological reservoirs, and ice systems, significantly alters the transport properties \citep{de1983hydrodynamic,   padilla1999effect, bekri2002dispersion, vanderborght2007review, cueto2008nonlocal}. Experimental, numerical and theoretical studies have attempted to derive relationships between multiphase flow properties and transport dynamics \citep{nutzmann2002estimation, zoia2010continuous, sahimi2012dispersion,   bromly2004non,jimenez2015pore, jimenez2017impact}. In transient multiphase flow regimes, capillary forces produce highly intermittent velocity fields \citep{roman2016particle} and can generate large-scale avalanches in the mobility of the fluid phases \cite{dougherty1998distribution, planet2009avalanches, tallakstad2009steady, zhao2016wettability}. Yet, how the intermittent nature of multiphase flows influences dispersion remains unknown.

\begin{figure}[t!]
\includegraphics[width=.46\textwidth]{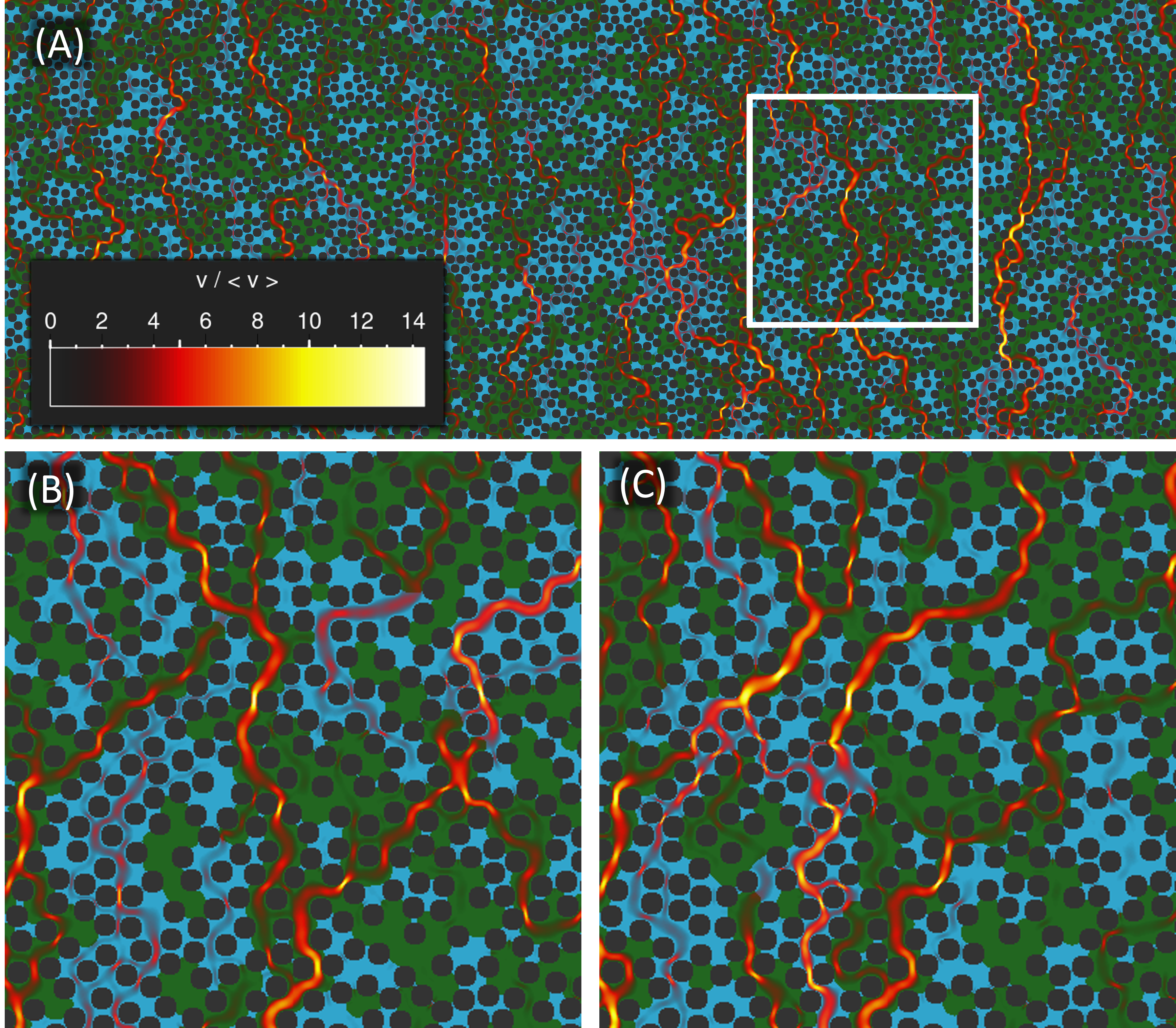}
\caption{Simulation of the flow of two immiscible fluids in a porous medium ($\mathrm{Bo}=0.18$). The flow is driven by a body force pointing in the vertical direction from the bottom to the top. The non-wetting (green) and wetting (blue) fluids flow between fixed cylindrical obstacles (black circles) forming the porous medium. The local fluid velocities are indicated by red to white colors overlaying the fluid phases. Panel A) shows the full system in which we have performed our simulations and panels B)-C) show zooms inside the white box of panel A) for two different time steps. Overall the flow is highly heterogeneous and intermittent, as illustrated by the temporal variations in the flow pathways between the panels B) and C).
}\label{fig:setup}
\end{figure}

Here we use extensive simulations of large two-dimensional porous systems based on the free energy Lattice-Boltzmann Method \cite{CONNINGTON2013601} to investigate dispersion in intermittent multiphase flows. While dispersion along the primary flow direction (longitudinal dispersion) is controlled primarily by the solid structure, transverse dispersion is enhanced by the highly intermittent flow field where local pathways transiently clog and unclog by the movement of the fluid phase boundaries (see Fig.\ \ref{fig:setup}). In the following, we argue that the relevant scale driving this enhanced dispersion follows from a non-linear function of the ratio between the surface tension and the body force driving the flow.

{\bf Methods} -- 
The simulations are performed in a cell of size 2048 by 768 lattice units filled with non-overlapping cylindrical obstacles (Fig.\ \ref{fig:setup}). The cylindrical obstacles have a fixed radius of $r=8$ lattice units. The position of the cylinders are determined by a Poisson process, which is repeated until the cell is filled. In this process, cylinders are only added if they have a minimum distance of $2r+1$ to existing center points. In the final state the porosity is 53\% (the fraction of space not covered by the cylinders). Note that the resulting porosity is large compared to natural porous media and that this is a requirement for maintaining good connectivity in two-dimensional systems \citep{deanna2013flow}. We expect the effect of intermittent multiphase flow on dispersion to be qualitatively similar in three-dimensions. 
 Our system has periodic boundary conditions in all directions, {\it i.e.}~when a fluid phase exits on one side, it reenters on the opposite side of the cell. The pore space is filled with two immiscible fluids, a wetting ($w$) and non-wetting ($nw$) fluid. The two fluid phases have a viscosity ratio $\nu_w/\nu_{nw}=0.57$ and a density ratio $\rho_w/\rho_{nw}=1.25$, respectively. The wetting angle is imposed to be 60 degrees and the non-dimensionalized surface tension $\gamma$ is varied in the interval $[0.005;0.040]$. In the case where $\gamma=0.0083$, our simulated system can be scaled to microfluidic device filled with a mixture of silicon oil and water. In addition to the multiphase simulations, we have computed the corresponding single-phase flow field satisfying the Stokes equations. All simulations are initiated in a state where the two phases are arranged in alternating stripes of width $2r$ transverse to the main flow direction. Dispersion is quantified once the distribution of phases is statistically stationary. Details about our numerical implementation and simulations are presented in the in the Supplementary Material S1 and our code is available at Ref.\ \footnote{\url{https://bitbucket.org/ulbm/felbm/}}.

We drive the flow by a body force density $\rho\mathbf g$ pointing along the shortest dimension of our cell, from the bottom to the top in Fig.~\ref{fig:setup}A. In the following, we express the driving force in terms of the dimensionless Bond number, which is the ratio between $g\equiv|\mathbf g|$ and the surface tension at the interface of the two fluid phases,
\begin{equation}
    \textrm{Bo}\equiv\frac{\ell^2 \rho  g}{\gamma},
    \label{eq:def_Bo}
\end{equation}
where $\rho=(\rho_w +\rho_{nw})/2$ is the mean of the two mass densities. The characteristic length $\ell$ is the average gap size between neighboring obstacles, where the neighbors are identified from a Voronoi tessellation. In our system, we have that $\ell=8.2$. Similarly, we describe the flow rate by the dimensionless Capillary number Ca$\equiv \mu_{w} \langle v\rangle/\gamma$, where $\mu_w$ is the dynamic viscosity of the wetting fluid and $\langle v\rangle $ is the average flow speed. In addition to varying the surface tension, we perform simulations with $|g|$ in the range $10^{-5}-10^{-3}$. All-in-all, our numerical simulations are performed for Bond numbers in the range $0.15$ -- $29.2$.

\begin{figure}[t!]
\includegraphics[width=.49\textwidth]{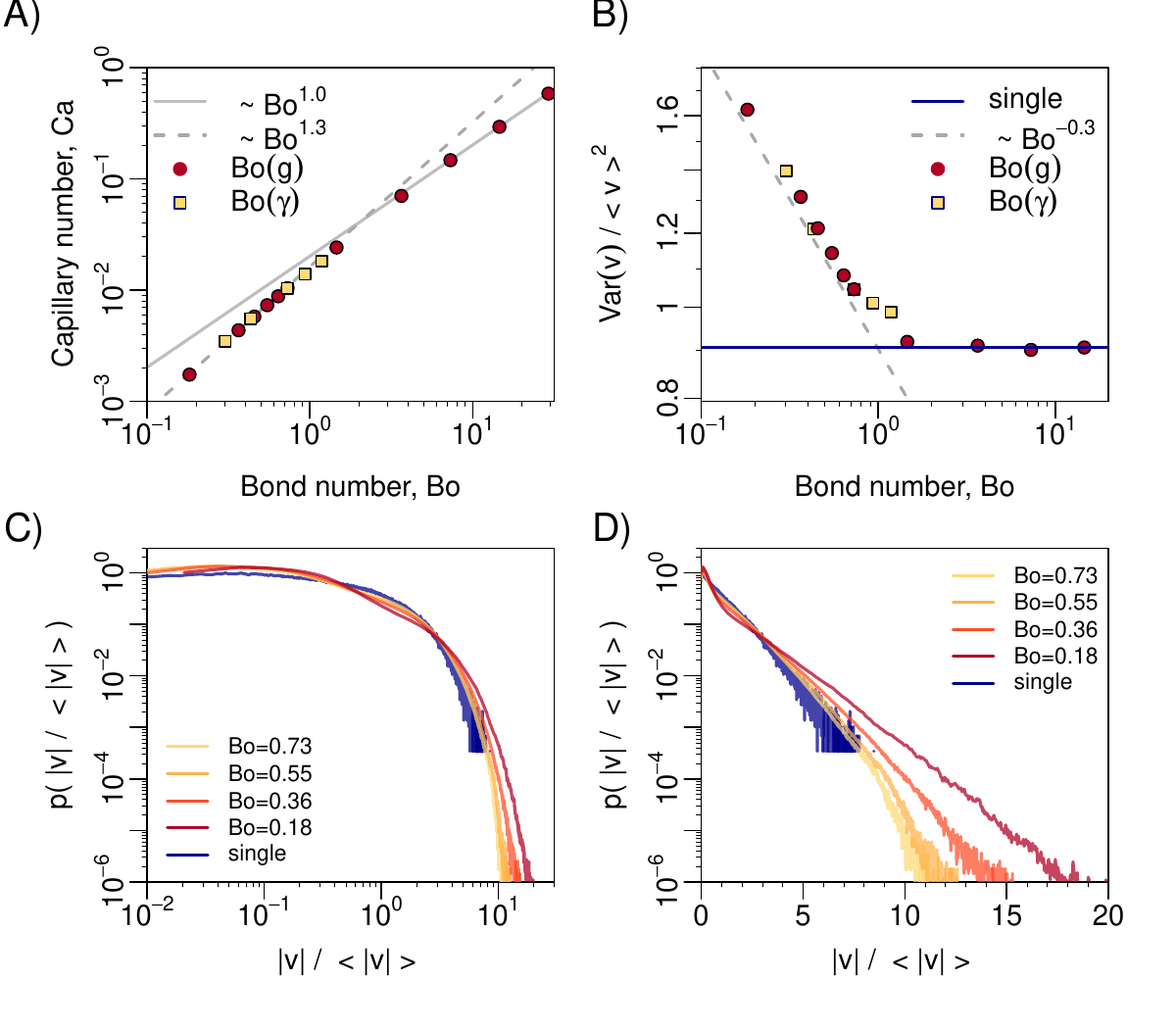}
\caption{Characteristics of the Eulerian velocity field for different Bond numbers. A), Capillary number (proportional to the average flow speed) as function of the Bond number. 
B) Variance of the flow speed relative to the average flow speed as a function of Bond number. C) and D) probability density function for the velocity magnitude on respectively double and semi-logarithmic coordinates.
}\label{fig:multi}
\end{figure}

{\bf Results} -- For large Bond numbers (Bo$>$1), our simulations are consistent with Darcy's law where the average flow rate is proportional to the applied gravitational force, leading to $Ca \sim Bo$. For small Bond numbers (Bo$<$1), the flow is significantly influenced by the surface tension and crosses over to a non-Darcian dynamic where Ca $\propto$ Bo$^{\beta}$ with $\beta=1.31$, consistent with experimental observations \citep{tallakstad2009steady,chevalier2015history, zhang2021quantification}, see Fig.~\ref{fig:multi}A and Supplementary Material S2 Movie.
For large Bond numbers, the velocity probability density function (pdf) is close to the single phase pdf and follows an exponential distribution (Fig.~\ref{fig:multi}D), as expected \citep{alim2017local}.
For low Bond numbers, we observe a broadening of the velocity distributions, which in the range of intermediate velocities become more power-law-like. This leads to an increase of the velocity variance relative to the mean velocity (Fig.~\ref{fig:multi}B)
All the distributions are characterized by a marked exponential cutoff at large values and a plateau at low values (Fig.~\ref{fig:multi}C-D). 

We inject passive tracers ($N=15000$) uniformly through the system in the initial state. At a time $t$, the particles reach a position
\begin{equation}\label{eq:partpos}
    \mathbf x^k(t)=\mathbf x^k(0)+\int_0^t \mathbf v\big(\mathbf x^k(t'),t'\big)\mbox{d}t',
\end{equation}
where $k$ is the index of a particle, $\mathbf v$ is local flow velocity and $\mathbf x^k(0)=\big(x^k_\perp(0),x^k_\parallel(0)\big)$ consists of respectively the longitudinal and transverse components. The particles passively follow the flow in the absence of molecular diffusion and are subject to the fluctuations induced by change in the spatial configuration of the fluid phases and the corresponding spatio-temporal heterogeneity of the flow velocity field. While we consider here the dispersion of purely advective particles, diffusion would be expected to further enhance transverse dispersion and induce a transition to Fickian dispersion for longitudinal dispersion.  
The statistics of the tracers is collected after the simulations reached a steady state (i.e.\ when there is no net-drift in the average flow rate). 

\begin{figure}[t!]
\includegraphics[width=.43\textwidth]{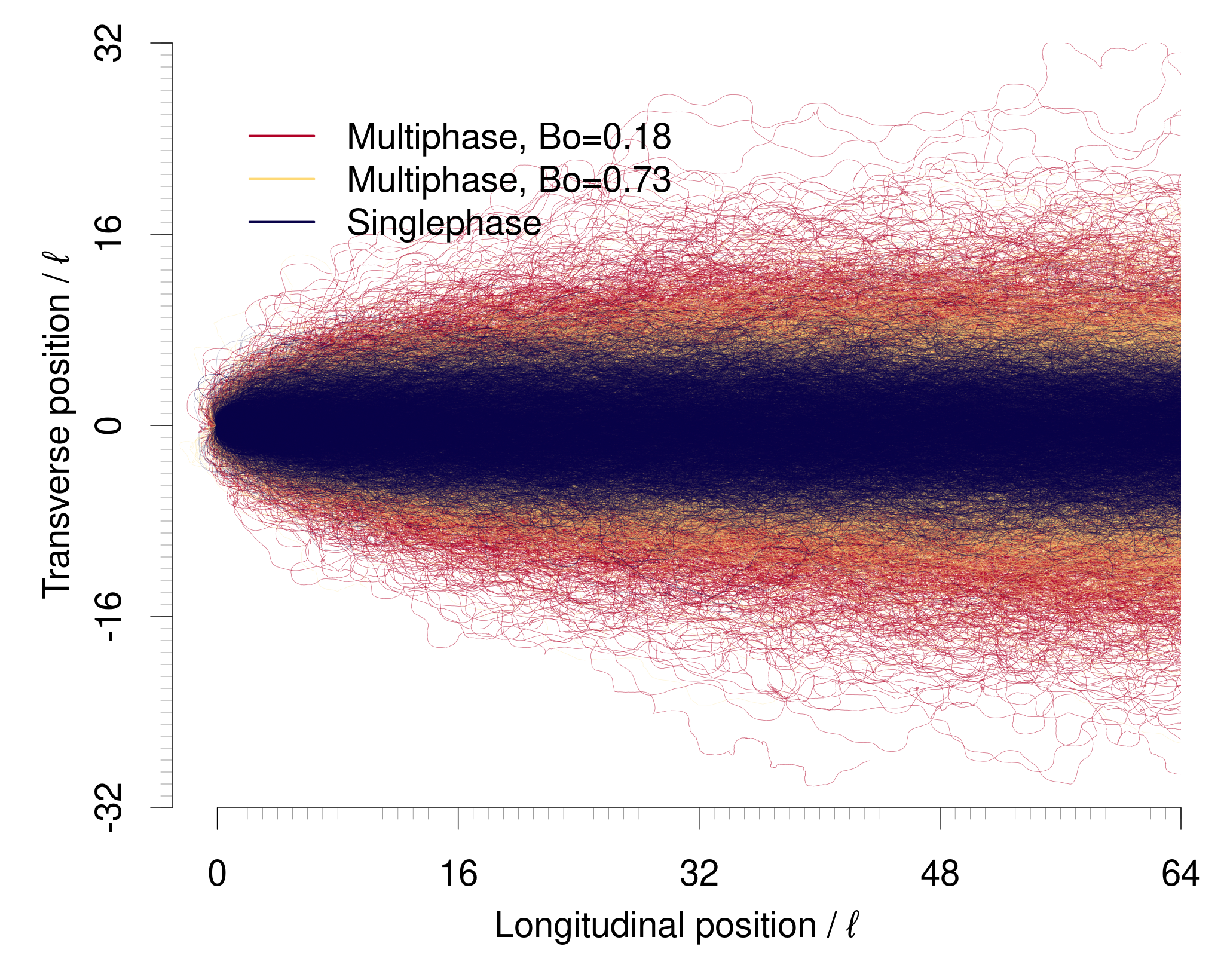}
\caption{Individual trajectories of passive tracers injected in three different flows, one multiphase flow at low speed (red, Bo=0.18), one multiphase flow at higher speed (yellow, Bo=0.73) and a single phase flow (blue). All the trajectories have been shifted so they start at the center of coordinates. For each of the three flows, we plot 4000 randomly selected trajectories.}
\label{fig:trajectories}
\end{figure}

In the single phase case, 
particle trajectories do not intersect, which implies that there is no asymptotic transverse dispersion in 2D. In the two-phase system, the flow can become temporarily blocked locally when the surface forces exceed the driving force of the fluid. In fact, repeated deactivation and reactivation of different flow paths (see the difference between the panels Fig.\ \ref{fig:setup}B-C and the Supplementary Material S2 Movie) allow particle trajectories to cross each other.
In Fig.\ \ref{fig:trajectories}, we show particle trajectories for a single phase and two multiphase simulations. In this figure the particle trajectories are shifted such that they all originate at the center of coordinates. 
 As the Bond number is lowered, the velocity fluctuations grow relative to the mean flow (see Fig.\ \ref{fig:multi}B) and lead to particle paths that look more jagged and wander further transversely (Fig.~\ref{fig:trajectories}). In the low Bond number limit, the local flow is highly intermittent and a few flow channels host larger proportions of the overall flow -- consistent with the experimental observations of critical avalanches in slow imbibition experiments \cite{dougherty1998distribution, planet2009avalanches}. In the following, we compute the dispersion from
the spatial variance of the position of passive tracers.
\begin{figure}[h]
\includegraphics[width=.48\textwidth]{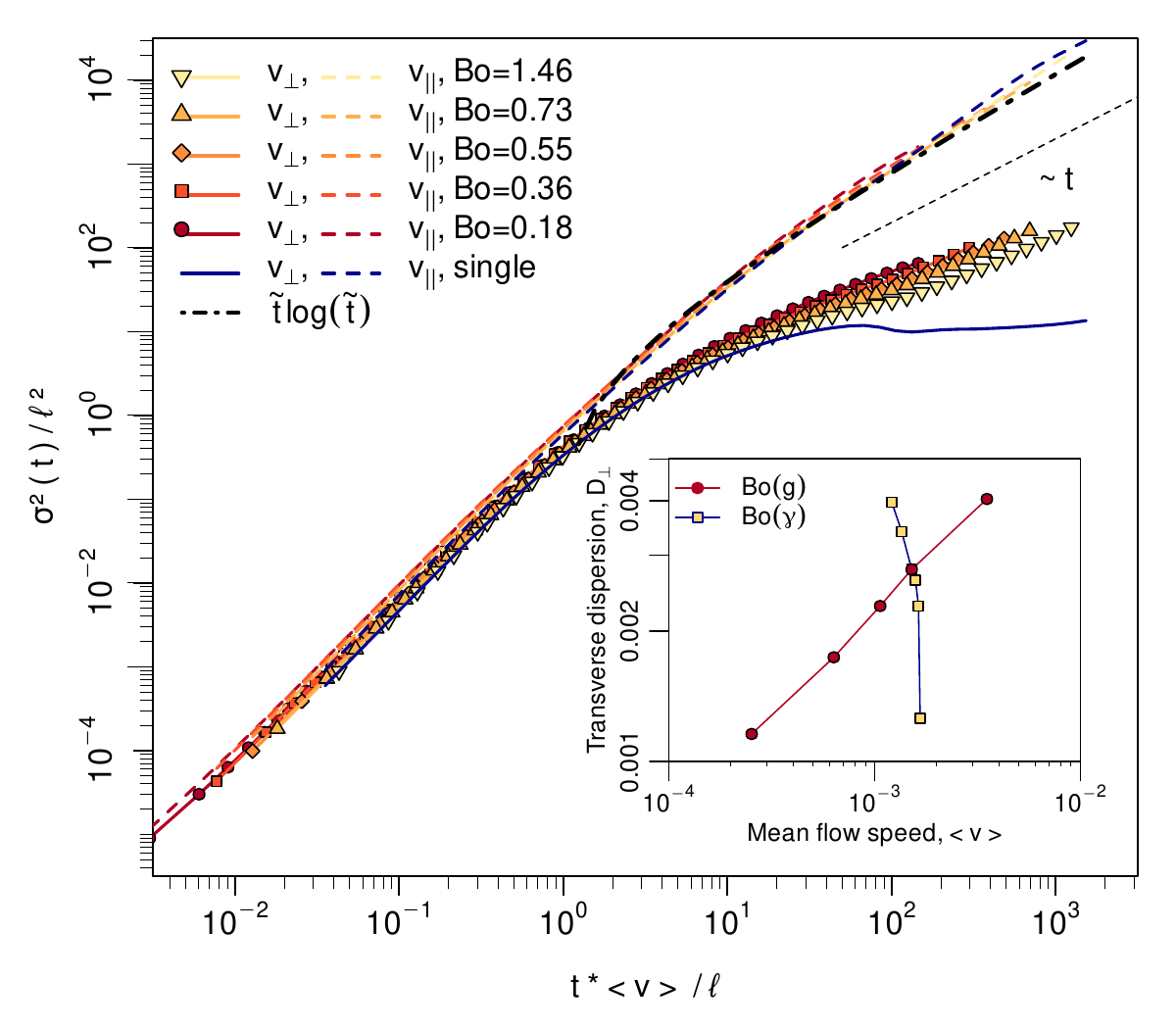}
\caption{Temporal evolution of longitudinal (${\parallel}$) and transverse (${\perp}$) spatial variances for single and multiphase flows with different Bond numbers using double logarithmic scales. 
The dashed lines represent the longitudinal particle dispersion along the flow direction. The solid lines with symbols represent the transverse dispersion. Colors yellow to red show five different Bond numbers for multiphase flow simulations (by varying the strength of the body force). The blue curves are computed from a single phase flow simulation and has a vanishing dispersion in the transverse direction.
The dashed straight line in black shows the Fickian scaling $\sigma^2 \sim t$, reached for transverse dispersion in two-phase flows. The inset shows the transverse dispersion coefficient computed from Eq.\ (\ref{sigma2}) as a function of average flow speed for the two cases where the body force, Bo($g$), respectively the surface tension, Bo($\gamma$), are varied. 
}
\label{fig:sigma}
\end{figure}

In Fig.\ \ref{fig:sigma}, we plot the diagonal components of the position variance tensor $\gv\sigma^2$ as function of time for different Bond numbers,
\begin{equation}\label{eq:sigma}
\gv\sigma^2(\Delta \tau)=\left\langle [ \mathbf x^k(\Delta \tau ) -\mathbf x^k(0) ]^T[ \mathbf x^k(\Delta \tau ) -\mathbf x^k(0)] \right\rangle_{k}.
\end{equation}
Here 
the average is performed over the ensemble of particles. We denote the diagonal components of the position variance tensor by $\sigma^2_\perp$ and $\sigma^2_\parallel$ in the transverse and longitudinal directions, respectively.

The longitudinal dispersion behaves similarly to that of a single-phase flow, which is apparent from Fig.\ \ref{fig:sigma} where the longitudinal variances for multiphase flow for different bond numbers all collapse on the single phase flow when normalizing time by the mean velocity.
Note that there is a slight shift between curves in the ballistic regime due to the non-linear dependency of the velocity variances with the Bond number (Fig. \ref{fig:multi}B). 
In the presence of a mean drift, the scaling of the spatial variance with time depends on the behavior of the velocity probability density function in the low velocity range \cite{dentz2016continuous}.
Broadly distributed velocity distributions can lead to non-Fickian dispersion, which may be quantified in the Continuous Time Random Walk framework (CTRW). Considering a power law scaling of the Eulerian velocity distribution $p(v)\sim v^{\alpha}$, 
the prediction of CTRW is that 
dispersion is Fickian for $\alpha<0$ and non-Fickian for $\alpha \geq 0$.
The almost uniform velocity distribution at low velocities (Fig.\ \ref{fig:multi}C-D), 
implies that the CTRW Lagrangian travel time distribution scales as $p(t)\sim t^{-3}$ and thus that dispersion is non-Fickian and follows \citep{dentz2016continuous},
\begin{equation}
\sigma_\parallel(t)\stackrel{t>\tau_a}{\approx} t/\tau_a \ln{t/\tau_a}.
\end{equation}
This behavior is verified in Fig.\ \ref{fig:sigma}, for times larger than the characteristic advection time, $\tau_a=\ell/\langle v \rangle$. Note that in multiphase flow systems composed of water and air, with much larger viscosity and density ratios, the emergence of trapped clusters and dead end zones leads to broader velocity distributions and strongly anomalous transport  
\citep{velasquez2021sharp}.

In contrast to the longitudinal behavior, the late time evolution of the transverse dispersion is Fickian, $\sigma^2\sim t$ (Fig. \ref{fig:sigma}). The transverse dispersion varies with the flow rate and is distinctly different from the vanishing dispersion of two-dimensional single phase flows (Fig. \ref{fig:trajectories}). The multiphase dispersion curves do not collapse on the single phase curve. In fact the dispersion fans out for the different Bond numbers at larger times. The dispersion coefficient follows from the asymptotic behavior of $\sigma^2_\perp$, 
\begin{equation}\label{sigma2}
D_\perp=  \lim_{\Delta\tau\rightarrow \infty} \left(\frac{\sigma^2_{\perp}(\Delta \tau)}{2\Delta\tau} \right).
\end{equation}
In the inset of Fig.\ \ref{fig:sigma}, we plot the transverse dispersion computed using Eq.\ (\ref{sigma2}) as function of the mean flow speed. The dispersion behaves differently when we vary the force driving the flow in comparison to when we vary the surface tension. In the latter case, we observe that an increase in surface tension enlarges the amplitude of the transverse motion without modifying significantly the overall flow speed. Achieving the same enhancement in dispersion by varying the driving force requires changing the mean velocity by one order of magnitude (see inset of Fig.\ \ref{fig:sigma}).
In the following, we propose a mechanistic model leading a scaling function for the dispersion that collapses the points in the figure onto a single curve.

\begin{figure}[t]
\includegraphics[width=.48\textwidth]{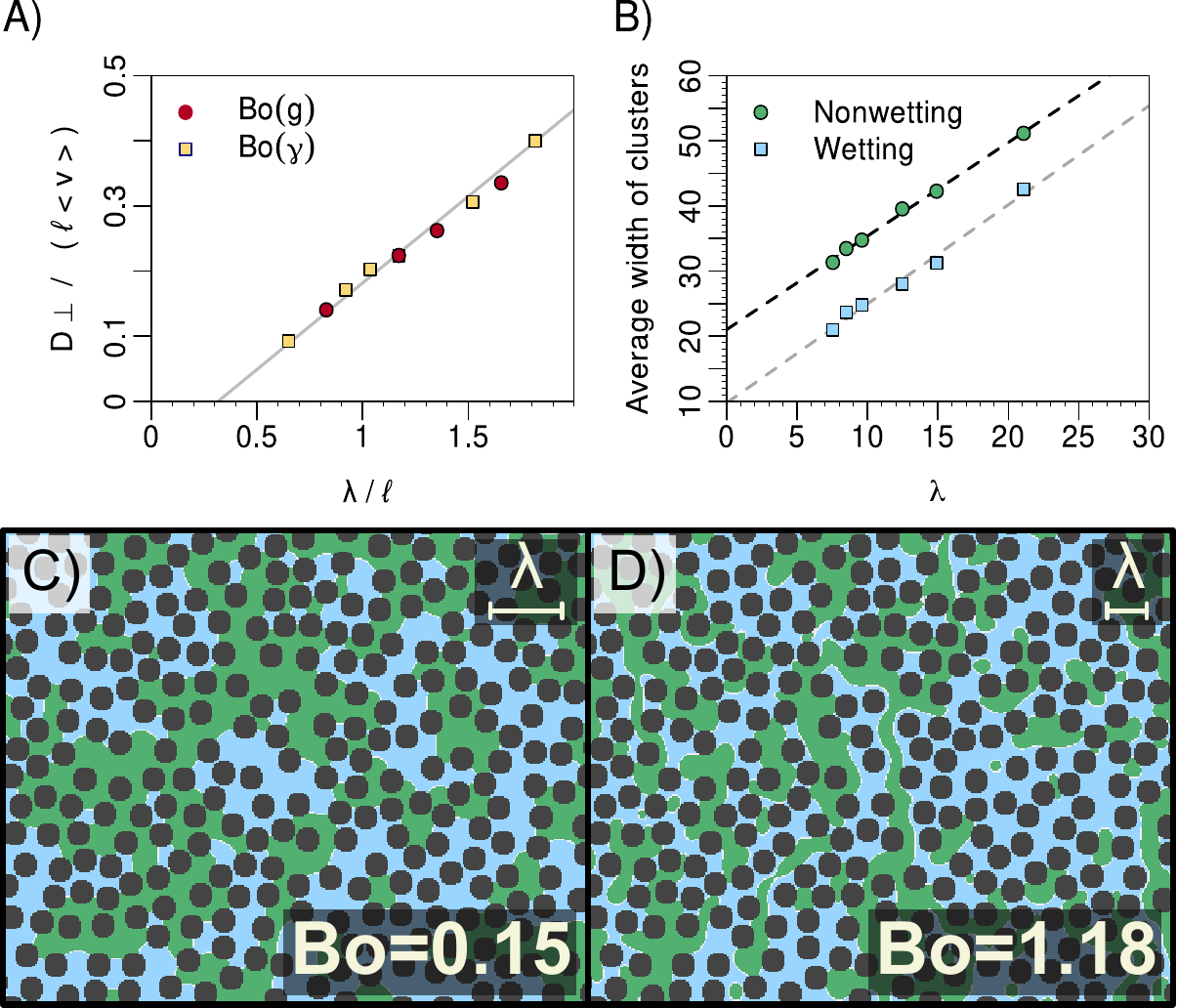}
\caption{A)  Transverse dispersion as a function of the effective length scale $\lambda$. The grey line in the background is a best linear fit (slope = 0.26$\pm 0.02$). B) Average transverse width of fluid clusters as a function of $\lambda$ for respectively the wetting (blue squares) and the non-wetting phases (green circles).
The grey lines are best linear fits to the points with a slope = 1.5$\pm 0.1$ for the wetting clusters and a slope = 1.4$\pm 0.1$ for the non-wetting clusters. Panels C) and D) show snapshots of two simulations performed at Bo=0.15 and Bo=1.18. The scale bar shows the corresponding length scale $\lambda$. For large Bond numbers, the domains of the wetting fluid become more fragmented by the non-wetting phase. }\label{fig:caplength}
\end{figure}

Because there is no mean drift in the lateral direction, the broad distribution of velocities does not impact the scaling of the transverse dispersion. For the type of persistent random walks performed by the passive tracers \citep{masoliver2017continuous}, the dispersion coefficient is proportional to the correlation length squared and divided by the correlation time $\tau$. The transverse motion is expected to correlate over a scale set by the typical cluster size $l_c$, hence
\begin{equation}\label{transverse_disp}
D_\perp \propto l_c^2/\tau=l_c \langle v \rangle,
\end{equation}
 with $\tau=l_c / \langle v \rangle$. The change in cluster size with Bond number is visible in Fig.\ \ref{fig:caplength}C-D where we present snapshots of simulations at two different Bond numbers. The cluster size increases when decreasing Bond number, which provides a mechanism for enhancing dispersion (Fig. \ref{fig:trajectories}).
 
As seen from  simulations 
(see Movie S2 in Supplementary Material), the size of wetting phase clusters (or domains) is limited by fragmentation due to localized drainage. We hypothesize that, for a given cluster size $l_c$, the probability of fragmentation depends on the probability that the cluster perimeter includes pores whose threshold capillary pressure for drainage is larger than the average pressure jumps across the cluster $\Delta p = l_c \rho g$. We further assume that the distribution of threshold capillary pressures for drainage follows an exponential distribution, $p(p_c)= \eta^{-1} e^{-p_c/\eta }$. The average threshold capillary pressure $\eta\propto\gamma / \ell$ is proportional to the surface tension divided by a characteristic length scale for a fluid interface moving through a pore throat, which again is proportional to the typical gap size $\ell$. The probability that the pressure difference in a given pore on the cluster perimeter is less than the threshold is $q=e^{- \Delta p /\eta}$.  The probability that all pores on the cluster perimeters are below the threshold is thus $q^n$, with $n\sim l_c/\ell$. The probability to maintain a cluster of size $l_c$ is therefore $p(l_c)\sim \exp{\left[-l_c^2 \rho g/( \eta \ell)\right]}=\exp{\left[-l_c^2 \rho g/\gamma\right]}$, which leads to an average cluster size, 

\begin{equation}\label{CL}
\langle l_c\rangle \sim \lambda \equiv \sqrt{\frac{\gamma}{\rho  g}}.
\end{equation}
In Fig.\ \ref{fig:caplength}B, we plot the average width of clusters (measured as the transverse extent) as function of $\lambda$, which confirms Eq.\ (\ref{CL}). Note that the theory presented here may be generalized to different scaling functions for cluster size that can develop in other multiphase flow regimes, with much larger viscosity or density ratios \citep{tallakstad2009steady, tallakstad2009steadyPRE}.

 From Eqs.\
 (\ref{transverse_disp}) and (\ref{CL}) we obtain the following relation for the transverse dispersion
\begin{equation}\label{scaling}
    D_\perp\propto \lambda \langle v \rangle=   \frac{\ell \langle v \rangle}{\sqrt{\textrm{Bo}}},
\end{equation}
which is confirmed by the data collapse in Fig.\ \ref{fig:caplength}A from the inset of Fig.\ \ref{fig:sigma}. 
The generality of Eq.\ (\ref{scaling}) is verified for variations in the surface tension and in the force driving the flow. We note that only by Eq.\ (\ref{scaling}) can we account for the dispersion when we vary the surface tension, since here the flow speed does not change much and yet we observe a significant change in the dispersion. Future work should explore to which extent our results hold when we consider systems with a different fluid saturation, different viscosity and density differences, and three-dimensions.

{\bf Concluding remarks} -- 
Temporal intermittency is a key property of multiphase flows in porous media that makes them fundamentally different from single phase flows. While longitudinal dispersion is not significantly affected by these temporal fluctuations, the repeated activation and deactivation of flow paths significantly enhances transverse dispersion with respect to single phase flow.
This phenomenon occurs in the low Bond number regime, where surface tension plays an important role with respect to gravitational forces, leading to highly fluctuating flows. We have shown that dispersion in such flows
is controlled by the length scale of fluid clusters, given by 
Eq.\ (\ref{CL}).
A direct consequence of our finding is that the transverse dispersion can be tuned by altering the ratio between the surface forces and the force driving the flow. This opens new perspectives for understanding, modelling and controlling transport properties in natural and engineered porous media systems. 

{\bf Acknowledgement} -- This study received funding from the Akademiaavtaalen between the University of Oslo and Equinor (project MODIFLOW to F. Renard).

\bibliographystyle{apsrev4-2}
\bibliography{references}
\end{document}